\documentclass[12pt]{article}
\usepackage{amssymb,latexsym}
\usepackage[mathscr]{eucal}

\newcommand{\ft}[2]{{\textstyle\frac{#1}{#2}}}

\newcommand{\g}{{\mathfrak g}}

\newcommand{\Z}{{\mathbb Z}}
\newcommand{\R}{{\mathbb R}}
\newcommand{\C}{{\mathbb C}}

\newcommand{\Lie}{{\mathcal L}}
\newcommand{\beq}{\begin{equation}}
\newcommand{\eeq}{\end{equation}}
\newcommand{\beqa}{\begin{eqnarray}}
\newcommand{\eeqa}{\end{eqnarray}} 
\newcommand{\tr}{\mbox{\rm Tr}} \newcommand{\uno}{\mbox{1
\kern-.59em {\rm l}}}

\newcommand{\tinyyoung}[1]{\mbox{\tiny\young(#1)}}
\usepackage[vcentermath,enableskew]{youngtab}
\newcommand{\nn}{\nonumber}
\begin{document}
\begin{titlepage}
\begin{flushright}
{ROM2F/2004/06}\\
\end{flushright}
\begin{center}

{\large \sc Multi instanton calculus on ALE spaces }\\

\vspace{0.2cm}
{\sc Francesco Fucito}\\
{\sl Dipartimento di Fisica, Universit\'a di Roma ``Tor Vergata'',
I.N.F.N. Sezione di Roma II\\
Via della Ricerca Scientifica, 00133 Roma, Italy}\\
{\sc  Jose F. Morales}\\
{\sl Dipartimento di Fisica, Universit\'a di Torino \\
 Laboratori Nazionali di Frascati,  \\
P.O. Box, 00044 Frascati, Rome, Italy }\\
and\\
{\sc Rubik Poghossian}\\
{\sl Yerevan Physics Institute\\
Alikhanian Br. st. 2, 375036 Yerevan, Armenia}\\
\end{center}
\vskip 0.5cm
\begin{center}
{\large \bf Abstract}
\end{center}
{ We study SYM gauge theories living on ALE spaces.
  Using localization formulae we compute the prepotential (and its
gravitational corrections) for $SU(N)$ supersymmetric ${\cal N}=2,
2^*$ gauge theories on ALE spaces of the $A_n$ type.
Furthermore we derive the Poincar\'{e} polynomial
 describing the homologies of the corresponding moduli spaces of self-dual
 gauge connections. From these results we extract
 the ${\cal N}=4$ partition function which is a
 modular form in agreement with the expectations of $SL(2,\Z)$
 duality. }

\par    \vfill
\end{titlepage}
\addtolength{\baselineskip}{0.3\baselineskip}

\tableofcontents

\setcounter{section}{0}
\section{Introduction}
In the past two years localization techniques have proved to be
very useful for the computation of non perturbative effects in
gauge theories \cite{Nekrasov:2002qd}-\cite{Nekrasov:2004vw}.
Besides being very powerful, these techniques are also very
flexible and can be applied in a variety of different contexts.
This work, in which we compute non perturbative effect (i.e. the
prepotential) for ALE manifolds of the $A_p$ type is an explicit
demonstration of this. The viewpoint that allows
for such unifying treatment is that of D-branes. In fact the bound state
of $k$ D(-1) and $n$ D3 branes describes, in the $\alpha^\prime\to
0$ limit, the moduli space, ${\cal M}$,  of gauge connections for
the supersymmetric Yang-Mills gauge theory (SYM) with sixteen
supercharges \cite{wttn,Douglas:1998uz,dhkmv}. This prototype
example is the starting point for further generalizations. The
presence of the D3 branes naturally suggests to consider the
original ten dimensional space $\R^{10}$ as $\R^4\otimes\R^6$ and
the original symmetry group $SO(10)$ as $SO(4)\times SO(6)\sim
SU(2)_L\times SU(2)_R\times SU(4)$. These spaces can then be
modded by some discrete $\Z_p$ group \cite{dmjm}. Modding the
first $\R^4$ by embedding the $\Z_p$ into one of the two $U(1)$'s
of $SU(2)_L\times SU(2)_R$, we get ALE spaces, while if we embed
$\Z_p$ into one of the two $SU(2)$'s in the maximal subalgebra of
$SU(4)$ we find certain quiver gauge theories. The latter case
will be the subject of a separate publication \cite{fmp}.

For some of the present authors, the interest in such models was
sparked by their relation with the AdS/CFT correspondence. It is
in fact well known that in the ${\cal N}=4$ SYM the space-time
geometry $AdS_5\times S^5$ is replicated in the moduli space of
gauge connections \cite{dhkmv}. The same holds true for the cases
with lower supersymmetries in which large $N$ geometries of the
type $AdS_5\times S^5/\Z_2$, $AdS_5\times S^3$, $AdS_5\times S^1$
and $AdS_5\times S^5/\Z_p$ \cite{gns,hkm,hk} were recovered. At
last these examples were reconsidered in \cite{Fucito:2001ha} from
the D brane viewpoint: according to the type of probe used to test
the space-time geometry, the $AdS_5\times S^1$ (pure ${\cal N}=2$
SYM), $AdS_5\times S^5/\Z_p$ (quiver case) and $AdS_5/\Z_p\times
S^5$ (ALE case) cases were recovered at finite $n$. The present
paper draws largely from this last reference and from
\cite{Bianchi:1996zj} in which SYM theories on ALE instantons were
studied as exact solutions of the four dimensional heterotic
string equations of motion with constant dilaton and zero torsion.
The techniques of the two last mentioned reference can be very
profitably incorporated into the scheme of the localization
described in \cite{Flume:2002az,Bruzzo:2002xf}.

For localization to happen the gauge
theories of our interest need to be deformed in a suitable.
For these four dimensional theories such deformation is
provided by a rigid rotation under the torus
$T_\epsilon=U(1)\times U(1)$ acting as $(z_1,z_2)\to
z_\epsilon=(e^{i\epsilon_1}z_1,e^{i\epsilon_2}z_2)$ where $z_1,
z_2\in\C^2$ are complex coordinates on the Euclidean space time.
Moreover the moduli space of gauge connections must be made compact
by introducing a regularization, $\zeta$.
The blow down ALE is defined via an orbifold projection with
$\Gamma$ inside $T_\epsilon$ \cite{Fucito:2001ha,Bianchi:1996zj}.
This implies in particular that fixed
points under $T_\epsilon$ are automatically invariant under
$\Gamma$ and therefore the analysis on $\R^4$ adapts easily to
the ALE case.
In addition localization provides us with a powerful tool to
compute the homologies of ${\cal M}$ as we will explain later
following \cite{nakajima}.

All of the results in the present paper are obtained for blown
down ALE spaces and in the limit of vanishing $\zeta$. Still our results
are valid for the full ALE due to the topological nature of the quantities
we compute \cite{nakajima,Nekrasov:2002qd,Flume:2002az}.

This is the plan of the paper: in section 2 we recall some details
of the solution of ${\cal N}=2, 2^*$ SYM theories on $\R^4$ that
will be important for the following. In this very section we also
briefly recall the construction of gauge connections on ALE
manifold which appeared in \cite{kronnak}. In section 3 we study
the cohomologies of the moduli spaces of self dual gauge
connections on ALE manifolds and compute the ${\cal N}=4$
partition function which turns out to be a modular form. This is a
strong check for the conjectured invariance (in the ``weak" sense
described in \cite{Vafa:1994tf}) of ${\cal N}=4$ under
electro-magnetic duality.
 Finally in section 4 we compute, for arbitrary values of the
winding number, the prepotential for a ${\cal N}=2, 2^*$ SYM
theory living on an ALE manifold. To make these results more
transparent we write down the lowest terms in the expansion of the
prepotential giving also the gravitational corrections.

\section{Preliminaries}
\setcounter{equation}{0}
\subsection{Localization on the ADHM manifold}
\label{gaugesec}  The ADHM construction can be seen as a way to
construct non flat hyperkh\"{a}ler manifolds of a given dimension
starting from completely flat manifolds of dimension greater than
the given one. The coordinates of the flat manifolds are organized
in the ADHM matrix $\Delta=a+bx$. Due to the symmetries of the
ADHM construction (we will later come back to this point at
greater length) we may choose the matrix $b$ so that it does not
contain any moduli. Then for the gauge group $SU(n)$, the matrix
$\Delta$ can be written as
\beq \label{salute1}
\Delta=a+bz=\pmatrix{J & I^\dagger\cr B_1 &
-B_2^\dagger\cr B_2 & B_1^\dagger}+\pmatrix{0 & 0\cr z_1 &
-\bar z_2\cr z_2 & \bar z_1}.
\eeq
Here $z_{1,2}$ (which in (\ref{salute1}) are meant to be multiplied by a unit
$[k]\times [k]$ matrix) parameterize the
position in the base space $\R^4$ while $m=\{
J,I^\dagger,B_1,B_2\}$ are coordinates on the flat
$4k^2+4kn$-dimensional hyperkh\"{a}ler manifold $M=
\R^{4k^2+4kn}$. More precisely $B_{1,2}$ and $J, I^\dagger$ are $[2k]\times
[2k]$ and $[n]\times [2k]$ dimensional matrices respectively. They
can be thought of as homomorphisms $B_{1,2}:V\times Q \to V $
and $J:V\to W\times \Lambda^2 Q$, $I:W\to V$.
$V, W$ are $[k]$ and $[n]$ dimensional spaces respectively.
$Q$ is the 2 dimensional chiral spin space.
$1\oplus\Lambda^2 Q\equiv 1\oplus Q\wedge Q$ is the antichiral spin space.
A self-dual field strength is built out of the ADHM connection
\beq A_\mu=\bar U(x)\partial_\mu U(x).
\label{gaugeconn}
\eeq
with $U$ a $[2k+n]\times [n]$ matrix satisfying $\bar\Delta U=0$.
Given the three complex structures $J^i_{ab}$ where $i=1,2,3$ and $a,
b=1,\ldots,{\rm dim}\, M$, we can build the 2-forms
$\omega^i=J^i_{ab}dm^a\wedge dm^b$. The real forms $\omega^i$
allow one to define a $(2,0)$ and a $(1,1)$ form
\beqa
\label{comforms}
\omega_{\C}&=& \tr\, dB_1\wedge dB_2+\tr\, dI\wedge dJ\ \ ,\nonumber\\
\omega_{\R}&=&\tr\, dB_1\wedge B_1^\dagger+\tr\, dB_2\wedge
dB_2^\dagger+ \tr\, dI\wedge dI^\dagger- \tr\, dJ^\dagger\wedge
dJ\ \ .
\eeqa
The ADHM data is invariant under
\beqa\label{uk}
B_1&\rightarrow& T_{\epsilon_1}T_\phi B_1T_\phi^{-1},\nonumber\\
B_2&\rightarrow& T_{\epsilon_2}T_\phi B_2T_\phi^{-1},\nonumber\\
I&\rightarrow& T_\phi I T_a^{-1},\nonumber\\
J&\rightarrow& T_{\epsilon}T_a JT_\phi^{-1} \eeqa with
$T_\phi=e^{i\phi}\in U(k)$,
$T_a=\rm{diag}(e^{ia_1},\ldots,e^{ia_n})\in SU(n)$ ,$ T_{
\epsilon_{1,2}}=e^{i\epsilon_{1,2}}\in U(1)^2$. The
transformations $U(n)\times  U(1)^2$ describe the gauge and
Lorentz symmetries of the theory while $U(k)$ parameterizes the
redundancy of the ADHM construction. Having fixed a basis
$\{e_a\}$ of $\g=U(k)$, the condition that the Lie derivative
annihilates the two forms $\omega^i$, $\Lie_{e_a}\omega^i=0$,
leads to conserved quantities. Using a complex notation for the
momenta $f^i_\xi=f^i_a e^a$, we get \beqa \label{mommap}
f_{\C}&=&[B_1,B_2]+IJ\ \,
\nonumber\\
f_{\R}&=&[B_1,B_1^\dagger]+[B_2,B_2^\dagger]+II^\dagger-J^\dagger
J.
\eeqa
The hypersurface $f_{\C}=f_{\R}=0$, modded out by the
$U(k)$ symmetries, is the ADHM moduli space ${\cal M}$.

In the non commutative case in which we allow
$[z_1,\bar z_1]=-\zeta/2, [z_2,\bar z_2]=-\zeta/2$
\cite{Nekrasov:1998ss}, the above condition becomes $f_{\C}=0,
f_{\R}=\zeta$. For $\zeta\neq 0$ the resulting space is compact.
 The localization formula described below is
valid only for compact spaces and therefore $\zeta\neq 0$ will be
always understood.

In the D-brane picture where instantons are
thought of as D(-1)-branes superposed over D3-branes, the homomorphisms are
realized by open strings connecting the $k$ D(-1)-branes and $n$
D3-branes and the ADHM constraints arise as D and F flatness
condition (with $\zeta$ the Fayet-Iliopoulos term) on the
effective $U(k)$ gauge theory living in the D-instanton.

Multi instanton corrections to correlators in four-dimensional
gauge theories can be cast in the form of an integral over the ADHM
manifold just described. Integrals over the ADHM manifold are in
general hard to handle, but for particular deformations of the
ADHM manifold they greatly simplify due to localizations.
The physical quantities can be extracted from the final result
after turning off the deformations in an suitable way.
Here we briefly describe the localization formula.

 Localization in the ADHM moduli space is based on the vector
field $\xi^*$, the \emph{fundamental vector field} associated with
the group element $\xi\in U(1)^{n-1}\times U(1)^2$,  that
generates the one-parameter group $e^{t \xi}$ of transformations
on ${\cal M}$. The vector field is parameterized by the elements
$T_a, T_{\epsilon_{1,2}}$ in the Cartan of the ADHM symmetry group
representing vevs and gravitational deformations. In the presence
of such deformation it is possible to see that the action of
${\cal N}=2$ SYM is invariant under a deformed BRST charge
operator $Q^*$, $\xi^*=1/2\{Q^*,Q^*\}$ , and it can interpreted as
a closed equivariant form \cite{Bruzzo:2002xf}. This BRST charge
implements the action of supersymmetry and of the symmetries
(\ref{uk}) on ${\cal M}$ and it can be identified with the
equivariant derivative $d_\xi =d+i_{\xi^*}$ with $i_{\xi^*}$ the
contraction using the vector field $\xi^*$. The localization
formula is now \beq \int_{\cal M}
\alpha(\xi)=(-2\pi)^{n/2}\sum_{x_0}{\alpha_0(\xi)(x_0)\over {\rm
det}^{1\over 2}\, {\cal L}_{x_0}} \label{locth} \eeq where
$\alpha(\xi)$ is an equivariant form ($e^{-S}$ in our case where
$S$ is the SYM action in ${\cal M}$) , $\alpha_0(\xi)$ its zero
degree part and ${\cal L}_{x_0}:T_{x_0} {\cal M}\to T_{x_0} {\cal
M}$ the map generated by the vector field $\xi^*$ evaluated at the
critical points $x_0$. See \cite{Bruzzo:2002xf} for a more
detailed explanation of the notations employed here. The critical
points are defined to be the fixed points $\xi^*(x_0)=0$ of the
vector field $\xi^*$ up to a diagonalizable $U(k)$ gauge
transformation given by
$T_\phi=\rm{diag}(e^{i\phi_1},\ldots,e^{i\phi_k})$,

>From the infinitesimal version of
(\ref{uk}) we find
\beqa \label{critical}
&&(\phi_{IJ}+\epsilon_\ell)\, B^\ell_{IJ}=0 \nn\\
&& (\phi_{I}-a_\alpha)\, I_{I\alpha}=0 \nn\\
&& (-\phi_{I}+a_\alpha+\epsilon)\, J_{\alpha I}=0 \eeqa with
$\epsilon=\epsilon_1+\epsilon_2$ and $\phi_{IJ}=\phi_I-\phi_J$.

The solutions of (\ref{critical}) can be put in one to one
correspondence with a set of $n$ Young tableaux $(Y_1,\ldots Y_n)$
with $k=\sum_\alpha k_\alpha$ boxes distributed between the
$Y_\alpha$'s. The boxes in a $Y_\alpha$ diagram are labelled
either by the instanton index $I_\alpha=1,\ldots,k_\alpha$ or by a
pair of integers $j_\alpha,i_\alpha$ denoting the vertical and
horizontal position respectively in the Young diagram. The
explicit solutions to (\ref{critical}) can then be written as
\beqa \label{solcritical} \phi_{I_\alpha}=\phi_{i_\alpha j_\alpha}
&=& a_{\alpha}+(j_\alpha-1)\epsilon_1+(i_\alpha-1)\epsilon_2 \eeqa
and $J=B_\ell=I=0$ except for the components $B_{1(i_\alpha
j_\alpha),(i_\alpha-1 j_\alpha)}$, $B_{2(i_\alpha
j_\alpha),(i_\alpha j_\alpha-1)},\,
I_{1,1},\,I_{k_\alpha+1,\alpha+1}$ $\alpha=1,\ldots,n$. At the
critical points the spaces $V, W$ become $T_\epsilon$ and $T_a$
modules allowing the decomposition \beqa\label{Vs}
V&=&\sum_{(i_\alpha,j_\alpha)\in Y_\alpha}
T_{a_\alpha}\,T_1^{-j_\alpha+1}
T_2^{-i_\alpha+1} \nonumber\\
W&=&\sum_{\alpha =1}^nT_{a_\alpha} \label{cinqueapp} \eeqa It is
then possible to compute the character as
\cite{Flume:2002az,Bruzzo:2002xf} \beqa \label{trace} \chi&=&
V^*\times V\times \left[T_1+T_2-T_1T_2-1\right] + W^*\times V
+V^*\times W\times T_1T_2\nonumber\\
&= &\sum_{\alpha, {\beta}}^n \sum_{s\in Y_j} \left(T_{a_{\alpha
{\beta}}} T_1^{-h_\beta(s)} T_2^{v_\alpha(s)+1}+ T_{ a_{
{\beta}\alpha}} T_1^{h_\beta(s)+1} T_2^{-v_\alpha(s)}\right) \eeqa
with $a_{\alpha {\beta}}=a_\alpha-a_{ \beta}$. $h_\beta(s)$
($v_\alpha(s)$) is the horizontal(vertical) distance from $s$ till
the right (top) end of the $\alpha (\beta)$ diagram, i.e. the
number of black (white) circles in Fig.1. See Appendix C of
\cite{Bruzzo:2002xf} for a more detailed explanation of the
computation and meaning of (\ref{trace}). The exponents in
(\ref{trace}) are the eigenvalues of the operator ${\cal L}_{x_0}$
which enters in our localization formula (\ref{locth}). Using
these eigenvalues and (\ref{locth}), the partition function of
${\cal N}=2$ SYM for winding number $k$ is
\cite{Flume:2002az,Bruzzo:2002xf} \beqa {\cal Z}_k=\sum_{x_0}
{1\over {\rm det}
 \hat{{\cal L}}_{x_0}} &=&\sum_{\{Y_\alpha; \sum_\alpha\vert
Y_\alpha\vert=k \}}
\prod_{\alpha,\beta=1}^n \prod_{s\in Y_{\alpha}}{1\over
E_{\alpha\beta}(s)(\epsilon-E_{\alpha\beta}(s))}
\label{generalsdet}
\eeqa
and
\beqa
E_{\alpha\beta}(s) &=&
a_{\alpha\beta}-\epsilon_1 h_\beta(s)+\epsilon_2(v_\alpha(s)+1)
\label{eeee}
\eeqa
${\cal N}=2^*$ SYM is obtained from pure ${\cal N}=4$ SYM giving
mass to the adjoint ${\cal N}=2$ hypermultiplet. In this latter
case the character is \cite{Bruzzo:2002xf} \beq \label{trace4}
\chi_m=(1-T_m^{-1})\chi \eeq where $\chi$ is the character of pure
${\cal N}=2$ SYM which was defined in (\ref{trace}). $T_m=e^{im}$
parameterizes the mass deformation. In fact, a mass deformation
can be introduced in a way similar to the equivariant deformation
$T_{\epsilon_{1,2}}$. While $T_{\epsilon_{1,2}}$ is embedded in
the $SU(2)\times SU(2)$ acting on Euclidean space time,
$T_m=e^{im}$ is a $SO(2)$ subgroup of the $SO(6)$ ${\cal
R}$-symmetry group which acts on the $\R^6$ space transverse to
the D3 branes we introduced before. In this case the partition
function for winding number $k$ is given by \cite{Bruzzo:2002xf}
\beq {\cal Z}_k =\sum_{\{Y_\lambda\}}
\prod_{\lambda,\tilde{\lambda}=1}^N \prod_{s\in
Y_\lambda}{(E_{\alpha\beta}(s)-m)(E_{\alpha\beta}(s)
-\epsilon+m)\over
E_{\alpha\beta}(s)(E_{\alpha\beta}(s)-\epsilon)}.
\label{generalsdetn=2*} \eeq
The
results for pure ${\cal N}=2$ SYM are easily obtained from
(\ref{generalsdetn=2*}) in the limit in which the mass of the
hypermultiplet decouples, i.e. $m\to\infty, m^4q=\Lambda$.

For simplicity we take
$\epsilon_2=-\epsilon_1=\hbar$.
It is convenient to introduce

the notation
\beqa
f(x)&=&{(x-m)(x+m)\over  x^2}
\nn\\
T_\alpha(x)&=&\prod_{\beta \neq \alpha}{(a_{\alpha
\beta}+x+m)(a_{\alpha \beta}+x-m)\over
 (a_{\alpha \beta}+x)^2}
\label{fst}
\eeqa
 The contributions coming from the first few tableaux can then be written as \cite{Bruzzo:2002xf}
\beqa
Z_{\tinyyoung{\hfil}}&=&\sum_\alpha f(\hbar) T_\alpha\label{zr41}\nn\\
Z_{\tinyyoung{\hfil},\tinyyoung{\hfil}}&=& \ft12
\sum_{\alpha\neq \beta}\, T_\alpha T_\beta \,
f(a_{\alpha\beta}+\hbar)f(a_{\alpha\beta}-\hbar){f(\hbar)^2 \over f(a_{\alpha\beta})^2}\label{zr42}\nn\\
Z_{\tinyyoung{\hfil}\tinyyoung{\hfil}}&=&
 \sum_\alpha
f(\hbar) f(2\hbar) T_\alpha T_\alpha(\hbar)\label{zr43}
\eeqa
 with $T_\alpha=T_\alpha(0)$ and $Z_{\tinyyoung{\hfil,\hfil}}$ given by $Z_{\tinyyoung{\hfil}\tinyyoung{\hfil}}$
 with $\hbar\to -\hbar$. The multi instanton partition function  $Z(q)=\sum_k Z_k q^k$
determines the prepotential ${\cal F}(q)$  via the relation \beq
{\cal F}(q)\equiv \lim_{\hbar\to 0}\,{\cal F}(q,\hbar)=
\lim_{\hbar\to 0}\,\hbar^2 ln \, Z(q) \label{prep} \eeq The
general function ${\cal F}(q,\hbar)$ encodes the gravitational
corrections to the ${\cal N}=2$ superpotential.

\subsection{The construction of Kronheimer and Nakajima}
\label{kronnakajima}
In this section we review the construction of gauge instantons on ALE spaces \cite{kronnak}.
ALE spaces can be obtained from the minimal resolution of orbifolds of the type $\R^4/\Gamma$.
In terms of the D-brane construction we discussed in the introduction,
the orbifold quotient is taken along the directions longitudinal
to the D3-brane system. These directions form a $\R^4$ space acted upon by
the Lorentz group $SO(4)\cong SU(2)_L \times SU(2)_R$.
$\Gamma$ is a discrete Kleinian subgroup of SU(2),
i.e. $\Gamma = \Z_p,D_N^*,O^*,T^*,I^*$. The explicit computations in the
next section will be carried out for the $\Z_p$ case only.
The recipe to get an ALE space is simple:
take a pair of $|\Gamma|\times|\Gamma|$ complex matrices $\alpha,\beta$
satisfying the $\Gamma$ invariance property
\beq
\gamma_v \pmatrix{ \alpha  \cr
 \beta \cr} \gamma_v^{-1} = \gamma_Q
\pmatrix{\alpha\cr \beta\cr}
\quad \gamma\in\Gamma
\label{gamminva}
\eeq
where $\gamma_v\in U(|\Gamma|)$ and $\gamma_Q\in SU(2)$ are matrices realizing
the element $\gamma\in\Gamma$ and $ (\alpha, \beta)$ transforming in the adjoint of
$U(|\Gamma|)$ and in the fundamental of $SU(2)$.
We then introduce a manifold $\Xi$ with coordinates $\Xi=(\alpha,\beta)$ of
real dimension
\beq
{\rm dim}\Xi = 2 \sum_{i,j=1}^{\vert \Gamma\vert} A_{ij} m_i m_j =
4 \sum_{i=1}^{\vert \Gamma\vert} (m_i)^2
= 4 | \Gamma |
\label{duenove}
\eeq
where $m_i$ are the dimensions of the irreducible representations, $R_i$ in the
decomposition $\Gamma=\sum_{i=1}^{\vert \Gamma\vert}m_i R_i$,
 $A_{ij} = 2 \delta_{ij} - \tilde C_{ij}$ with $\tilde C_{ij}$ the
extended Cartan matrix connected to $\Gamma$. See \cite{Bianchi:1996zj} for an exhaustive
description of these points. Starting from the datum $\Xi$, it is possible to build a
two form analogous to (\ref{comforms}), invariant under $U(\vert \Gamma\vert)$ transformations.
This invariance leads to two conserved quantities and to the constraints
\beqa
[\alpha,\beta] &=& \zeta_\C \cr
[\alpha,\alpha^\dagger]+[\beta,\beta^\dagger]&=&\zeta_\R
\label{duetredicibb}
\eeqa
where $\zeta\in \R^3\otimes{\cal Z}^*$, with
${\cal Z}^*$ the dual to the center of the Lie algebra of
$G=\otimes_{i=1}^{p-1}U(m_i)$,\footnote{$G$ is obtained from $U(|\Gamma|)$
keeping in account the previous decomposition of $\Gamma$.
In its decomposition we have omitted $U(m_0)$ whose action is trivial.} with
$\sum_{i=1}^{p-1} \zeta_i=0$.
Taking the quotient by $G$ we finally get a manifold of dimension
${\rm dim}\, X_{\zeta}=\dim\,\Xi-4\dim\,G =4\vert \Gamma\vert-4 (\vert \Gamma\vert -1)= 4$.

On these spaces it is then possible to extend the ADHM construction of the previous subsection.
With respect to what we already said, there
are two further requests which need to be satisfied
\begin{itemize}
\item The matrix $\Delta$ in
(\ref{salute1}) needs to be invariant under the action of $\Gamma$.
The projection is acting on the Lorentz indices as $\gamma_Q\in T_\epsilon$
and on Chan-Paton indices $V,W$ as $\gamma_v,\gamma_w$.
\item
The instanton solution is classified
by both the first and second Chern classes.
To properly define them we introduce a tautological bundle ${\cal T}$
with fiber the regular representation of $\Gamma$ and base the
ALE space itself. ${\cal T}$ keeps in account the fact that parallel
transporting a section of the bundle at infinity gives a holonomy
due to the non trivial topology of the base space.
\end{itemize}
Under the action of $\Gamma$ this tautological
bundle admits a decomposition ${\cal T}=\sum_q {\cal T}_q \otimes R_q$
with $R_q$ ($q=0,1,....p-1$) the irreducible representations of
$\Gamma$. The first Chern Class $c_1({\cal T}_q)$ of the ${\cal T}_q$
bundles, $q\neq 0$
( $c_1({\cal T}_0)=0$ ), forms a basis of the second cohomology group.
Under $\Gamma$ we get the decompositions
\beqa
V&=&\sum_q V_q\otimes R_q,\quad\quad W=\sum_q W_q\otimes R_q
\quad\quad Q=Q_1+Q_2\nn\\
\gamma_{v,w}R_q &=& e^{2\pi i q\over p}\, R_q
\quad\quad~~~~ \gamma_{Q}\, Q_1 = e^{2\pi i \over p}\, Q_1
\quad\quad ~~~\gamma_{Q}\, Q_2 = e^{-{2\pi i \over p}}\, Q_2
\label{znp}
\eeqa
The winding number $k=\sum_q k_q$ and the rank of the bundle
$n=\sum_q n_q$ are given in terms of the integers
$k_q={\rm dim} V_q,  n_q={\rm dim}W_q$.\footnote{The reader should
not confuse this decomposition with the partitions leading to the
Young tableaux we introduce in the previous subsection. The different
notation emphasizes this fact.}

 The decompositions ({\ref{znp}) determines that of the ADHM variables $I, J, B_l$. In
 particular the $\Gamma$-invariant components are
\beqa
 m_\Gamma=\{ I_{I_q\alpha_q},J_{\alpha_q I_q}, B_{1(I_{q+1} J_q)}, B_{2(I_{q-1} J_q)}  \}
\label{invale}
\eeqa
A similar result can be obtained for the fermions in the theory \cite{Fucito:2001ha}.
Notice that supersymmetry
is preserved by this projection since $\Gamma$ acts in the same way
on the different components of a given supermultiplet.

The moduli space of multi instanton solutions
and ADHM constraints can be described as before
through (\ref{salute1}), (\ref{mommap}) written in terms of the
invariant components (\ref{invale}).
In particular the dimension of the moduli space is
given by the total number of components in (\ref{invale}),
minus the number of ADHM constraints (given by
$3k_q^2$ for each $q$),
minus the dimension of $\prod_q U(k_q)$
\beq\label{dimmod}
{\rm dim}\, {\cal M} = 4\sum_q (k_q n_q+
\frac{1}{2}k_{q+1} k_q+\frac{1}{2}k_{q-1} k_q-k_q^2)
\eeq
The decomposition properties can be used to relate the Chern character of the
instanton bundle
to the Chern characters of the individual bundles
${\cal T}_q$.
The instanton bundle
is specified then by giving the first and second Chern class
\beqa
c_1&=&\sum_q \left( n_q-2k_q +k_{q+1}+k_{q-1} \right)
c_1({\cal T}_q)\nonumber\\
c_2&=&\sum_q \left( n_q-2k_q +k_{q+1}+k_{q-1} \right) c_2({\cal
T}_q)+\frac{k}{|\Gamma|}\label{c2} \eeqa
The restriction to the
interesting case of instanton solutions with vanishing first Chern
class imposes, as we will see, strong constraints on the allowed
instanton configurations $\{ k_q \}$ for a given partition $\{ n_q
\}$. In particular an instanton solution with vanishing first
Chern class is given by
\beq n_q-2k_q+k_{q+1}+k_{q-1}=0 \quad\quad
{\rm for}~~~q>0. \label{c10}
\eeq
This is a highly non trivial
constraint on the allowed values of $(k_q,n_q)$. Notice that only
in this case the instanton number defined as $k/|\Gamma|=k/p$
coincides with the second Chern class.

In \cite{Bianchi:1996zj} the reader will  be able to find detailed
discussions of various cases. Here we simply recall the results in the
simplest context: the $SU(2)$ gauge bundle on the Eguchi-Hanson
blown down space $\R^4/\Z_2$. Solutions to (\ref{c10}) in this case
are given by either $\vec{n}=(2,0), \vec{k}=(k,k)$ or
$\vec{n}=(0,2), \vec{k}=(k-1,k)$. They lead
to instanton solutions with integer and half-integer second Chern
class respectively, as can be easily seen from (\ref{c2}).
The dimension of the multi instanton moduli space can be read from
(\ref{dimmod}) and
turns out to be respectively equal to $8k$ and $8k-4$,
in agreement with  \cite{Bianchi:1996zj}.
 In the next section we derive again these results in a more general perspective
 and derive the corresponding ${\cal N}=2$ prepotential describing the
 low energy physics in the ALE.

\section{Cohomology of moduli spaces}
\setcounter{equation}{0}

 Localization is a powerful tool in the study of
cohomology. Here we apply this techniques to various instanton
moduli spaces. The basic idea is to associate a perfect Morse
function, i.e. a function $f(M)$ such that the part of the
manifold $M$ satisfying $f\le c$ for arbitrary positive $c$ is
compact, to a given momentum map defined by the action of a group
$G$ on $M$. Non-trivial p-cycles in $M$ are then in one-to-one
correspondence to the number of critical points $\partial_i
f(x_0)=0$ with $p$-negative eigenvalues for the Hessian
$\partial_{i}\partial_j f(x_0)$\cite{botttu}.

\subsection{$U(n)$ gauge theory on $\R^4$}

Let us start by considering the Poincar\'{e} polynomial
for the moduli space of $U(n)$ gauge connections of
winding number $k$, ${\cal M}_k^n $:
$P_t({\cal M}_k^n)=\sum_{n\ge 0}t^n{\rm dim}H^n({\cal M}_k^n)$\footnote{The case of $SU(n)$ is given
by choosing $\sum_{\alpha=1}^n a_\alpha=0$. This corresponds to factorize the center
of mass of the brane system. For the computations we carry on in this paper this
difference is immaterial.}.
First we observe that the momentum map obtained by acting on (\ref{comforms})
with the Lie derivative with respect to the action of $T^2_\epsilon$ and $T_a$
\beq
M(B_l,I,J)=\sum_{\alpha =1}^n\bigg\{\sum_{l=1}^2\sum_{i,j=1}^{k_\alpha}\epsilon_l\vert
(B_l)_{ij}\vert^2+\sum_{i=1}^{k_\alpha}(a_\alpha\vert I_{i\alpha}\vert^2+
(\epsilon_1+\epsilon_2-a_\alpha)\vert J_{\alpha i}\vert^2)\bigg\}
\label{morsefunc}
\eeq
is a perfect Morse function \cite{nakajima}.
In fact choosing $\epsilon_1\gg a_n\gg\ldots\gg a_1\gg \epsilon_2>0$ one fulfills
the condition that the part of ${\cal M}$ satisfying ${\cal M}\le c$ for
arbitrary positive $c$ is compact. From the point of view of the brane system,
the above choice means that the D(-1) instantons
(whose position in the transverse space is given by (\ref{solcritical}))
are far enough ($\epsilon_1\gg a_\alpha$) to feel
the whole $U(n)$ stack of D3 branes. This choice also insures that the D(-1) instantons never sit
on top of each other to avoid a singularity of the moduli space.

The Betti's numbers $b_{2q}={\rm dim}H^{2q}({\cal M})$ (remember we are discussing a complex case)
are given by the sum of all those Young tableaux $Y_\alpha$ with $q$
negative eigenvalues. Examining the character $\chi$ of (\ref{trace}) we find the negative eigenvalues of
the Hessian. They satisfy the conditions \cite{nakajima}
\begin{itemize}
\item $h_\beta(s)> 0$
\item $h_\beta (s)=0$ but $\alpha <   \beta$
\end{itemize}
The case $h_\beta(s)=0, \alpha= \beta, 1+v_\alpha(s)<0$, never happens.





At each row there is a single box with $h(s)=0$ and therefore the number of boxes
in $Y_\alpha$ with $h(s)=0$ is the number of rows $l_\alpha$.
The total number of negative eigenvalues in the Young tableaux pair $(Y_\alpha, Y_\beta)$ is
then given by
\beq
\sum_{\alpha, \beta}(k_\alpha-l_\alpha)+\sum_{\alpha< \beta}\,l_\alpha
=nk+\sum_{\alpha =1}^n(\alpha -1-n)l_\alpha.
\label{negeigen}
\eeq
$k_\alpha$ is the number of boxes in the $\alpha$-th diagram.
The generating function for the Poincar\'{e} polynomials is then
\beqa
\sum_{k=0}^\infty\, q^k\, P_t({\cal M}_k^n)&=&
\sum_{k=0}^\infty\sum_{\{Y_\alpha; \vert Y_\alpha\vert =k\}}q^kt^{2(nk+\sum_{\alpha}(\alpha -1-n)l_\alpha)}\nn\\
&=&\prod_{\alpha =1}^n\sum_{n_1,n_2,\ldots}\prod_{m=1}^\infty\bigg(q^mt^{2(n(m-1)+\alpha -1)}\bigg)^{n_m}\nn\\
&=&
\prod_{\alpha =1}^n\prod_{m=1}^\infty{1\over 1-q^mt^{2(n(m-1)+\alpha-1)}}.
\label{genfunc}
\eeqa
In writing this we use the standard trick to rewrite $k=\sum_\alpha \sum_{m=1}^{l_\alpha}\, {m n^\alpha_m}$,
$l_\alpha=\sum_{m=1}^{l_\alpha}\, 1$ and exchange the constrained sums with unconstrained
ones over the $n_1^\alpha,n_2^\alpha,\ldots$. $n_i^\alpha$ is the number of times the integer $i$
appears in the partition of $k_\alpha$.

A different choice of the relation between the parameters $\epsilon_{1,2}$ and the v.e.v.'s
$a_\alpha$ leads to a different physical situation.
Choosing $a_n\gg\ldots\gg a_1\gg \epsilon_1\gg\epsilon_2>0$ corresponds to D(-1) instantons living
in well separated D3 branes and the Poincar\'{e} polynomial reduces to the one of the case $U(1)^n$.

\subsection{Instantons on $\R^4/\Z_p$}

Next we consider instanton on $\R^4/\Gamma$ with  $\Gamma=\Z_p$.
The homologies of this space are the same of those for the manifold
in which the conical singularity is simply resolved.
The orbifold action is chosen  to be
\beq
a_\alpha\to a_\alpha+{2\pi  q_\alpha \over  p}, \quad\quad \epsilon_1\to \epsilon_1+
{2\pi \over p},\quad\quad \epsilon_2\to \epsilon_2-{2\pi \over p}.
\label{zn}
\eeq
where $q_\alpha,\quad \alpha=1,\ldots,n$ can take integer values between $0$ and $p-1$,
specifying the representations under which
the subspace $W_\alpha\in W$ transform. In particular the integers $n_q$ characterizing
the unbroken gauge group $\prod_q U(n_q)$ are given by the number of times that the
$q^{\rm th}$-representation appears in $W$, i.e. $n_q=\sum_\alpha \delta_{q,q_\alpha}$.
Similarly the ADHM auxiliary group $U(k)$ breaks into $\prod_q U(k_q)$ with $k_q$ the number of
instantons transforming in the representation $R_q$, i.e. $k_q={\rm dim} V_q$.
According to (\ref{zn}) the instanton associated to the box $(i_\alpha,j_\alpha)$ in the
tableaux $Y_\alpha$ transforms in the representation $R_{q_\alpha+i-j}$.
Take for example the following tableaux for a $U(4)$ gauge theory on $\R^4/\Z_3$
\beq
\tinyyoung{2,012,1201,2012,0120}\quad\quad \tinyyoung{1,20}
\quad\quad \buildrel 1 \over\bullet\quad\quad \buildrel 2 \over\bullet
\label{diagr}
\eeq
The number in the box in the bottom left position in the Young tableaux
in (\ref{diagr}) gives the representation, $R_q$, in which each D3 brane transforms.
The bullet stands for an empty Young tableaux which anyway transforms
under $\Z_3$.\footnote{In the example of (\ref{diagr}) this means that there are
no D(-1) branes superposed to the third and fourth D3 brane.}
>From (\ref{zn}) and (\ref{diagr}) we then infer $q_1=0,q_2=2,q_3=1, q_4=2$ i.e. $n=(1,1,2)$.
The number of instantons in each representation follows then by counting the number of
boxes in the Young tableaux labelled by the same integers $q_\alpha$.
For the example of (\ref{diagr}) we get $k=(6,6,7)$.
The multi instanton moduli space splits into disconnected pieces
${\cal M}_{(k_q,n_q)}$ specified by the set of integers $(n_q,k_q)$. As we will see
these components are simply connected, i.e. for each component we get $b_0=1$,
and support in general non-trivial cohomologies.

It is important to notice that $\Gamma$ given by (\ref{zn}) belongs to the ADHM symmetry group
$G=U(1)^2\otimes U(1)^{n}$ used for localization.
This implies that a fixed point under $G$ is automatically
invariant under $\Gamma$ and therefore
critical points on the ALE instanton moduli space
are given again in terms of $n$-sets of Young-tableaux with a total number of $k$ boxes.
Yet the contribution of each diagram to the determinant ${\rm det} {\cal L}_{x_0}$
in (\ref{locth}) will be substantially different since only $\Gamma$-invariant boxes are
now contributing. The same holds for the cohomology where the number of $\Gamma$-invariant
negative eigenvalues of the Hessian at a given critical point will be in general smaller than
the number of eigenvalues in the related flat space.

In addition the $\Gamma$-invariant analog of (\ref{trace}) is
\beq
\label{traceG}
\chi_\Gamma=\sum_{q} \left[ V_q^*\, V_{q+1}+
V_{q+1}^*\, V_q- V_q^*\, V_{q}\Lambda^2 Q-V_q^*\, V_{q} + W_q^*\,
V_q +V_q^*\, W_q\, \, \Lambda^2 Q \right]
\eeq
with $\Lambda^2 Q=Q_1Q_2$. The character (\ref{traceG}) is given by
\beqa
\chi_\Gamma &= &\sum_{\alpha,\beta}^n \sum_{s\in
Y_\alpha} \left(T_{a_{\alpha\beta}} T_1^{-h_\beta(s)} T_2^{v_\alpha(s)+1}+ T_{
a_{\beta\alpha}} T_1^{h_\beta(s)+1}
T_2^{-v_\alpha(s)}\right)\delta_{h_\beta(s)+v_\alpha(s)+1,q_\alpha-q_\beta}
\label{tracef}
\eeqa
As in the case of $\R^4$ we associate a non-trivial element of the
$2m$-cohomology group of moduli space of instanton in $\R^4/\Z_p$ to
each Young tableaux with $m$ $\Gamma$-invariant negative eigenvalues in (\ref{tracef}),
i.e. the number of boxes with $v_\alpha(s)+h_\beta(s)+1=q_\alpha-q_\beta~{\rm mod}~p$ and $h(s)>0$.

As an illustration consider $U(1)$ instantons on  $R^4/\Z_3$ with $n=(1,0,0)$
\beqa
&& k=(2,2,2): \quad
\tinyyoung{\bullet\hfil\hfil\bullet\hfil\hfil}
 +\tinyyoung{\hfil,\bullet\hfil\bullet\hfil\hfil}+
 \tinyyoung{\hfil,\hfil,\bullet\bullet\hfil\hfil}+
  \tinyyoung{\bullet\hfil\hfil,\hfil\bullet\hfil}+
   \tinyyoung{\hfil,\bullet\hfil,\hfil\bullet\hfil}+
   \tinyyoung{\hfil,\hfil,\times,\bullet\hfil\hfil}+
   \tinyyoung{\hfil\hfil,\bullet\hfil,\hfil\times}+
   \tinyyoung{\hfil,\hfil,\times,\hfil,\bullet\hfil}+
   \tinyyoung{\hfil,\hfil,\times,\hfil,\hfil,\times} \nn\\
&& k=(3,1,2):  \quad
 \tinyyoung{\hfil\hfil,\hfil\hfil\hfil\hfil}  \nn\\
&& k=(3,1,2): \quad
 \tinyyoung{\hfil,\hfil,\hfil\hfil,\hfil\hfil}
\label{diagr2}
 \eeqa
 Bullets and crosses stand for $\Gamma$-invariant boxes with $h(s)>0$ and $h(s)=0$
 respectively.
Counting the number of tableaux with a fixed number of bullets in (\ref{diagr2}) we find
 that the $k=6$ $U(1)$ instanton moduli space with $n=(1,0,0)$ on $R^4/\Z_3$ splits into three
simply connected ($b_0=1$) pieces which contribute to the Poincar\'{e} polynomial as
\beqa
q^6\,r^{2 R_0+2R_1+2R_2}(1+3 t^2+5 t^4) :&\quad &  b_0=1\quad b_2=3 \quad b_4=5 \nn\\
q^6\,r^{3 R_0+R_1+2R_2}   :&\quad &
 b_0=1  \nn\\
q^6\,r^{3 R_0+2R_1+R_2 } :&\quad &
 b_0=1
\eeqa
with $q=e^{2\pi i \tau}$.
The factor $r^{k_q R_q}$ describes the $\Gamma$-content.

{\bf $U(1)$ instantons}

Here we consider arbitrary instanton configurations for the $U(1)$ case.
For concreteness we take $n=(1,0,\ldots,0)$ and compute the Poincar\'{e} polynomial
as a series
$\sum_{m,k_q}\, b_{m}({\cal M}_{k_q})\,q^{\sum_q k_q} \,
t^m\,r^{k_q R_q}$.
  Let us start by considering $R^4/\Z_2$.
 The Poincar\'{e} polynomial of multi instanton moduli space on $R^4/\Z_2$ follows from that on
$R^4$ after decomposition in classes $k_0 R_0+k_1 R_1$
\beqa Y
&=&\bullet
+\tinyyoung{\hfil}+\left(\tinyyoung{\bullet\hfil}+\tinyyoung{\hfil,\times}\right)
+\left(\tinyyoung{\hfil\bullet\hfil}+\tinyyoung{\hfil,\hfil\hfil}+\tinyyoung{\hfil,\times,\hfil}\right)
+\left(\tinyyoung{\bullet\hfil\bullet\hfil}+\tinyyoung{\hfil,\bullet\bullet\hfil}
+\tinyyoung{\bullet\hfil,\hfil\times}
+\tinyyoung{\hfil,\times,\bullet\hfil}+\tinyyoung{\hfil,\times,\hfil,\times}
\right)+\ldots\nn\\
&=& 1+ q \,r^{R_0}+(1+t^2) q^2\, r^{R_0+R_1}+q^3 (
r^{2R_0+R_1}(1+t^2)+
r^{R_0+2R_1})+\nn\\
&&+ q^4 r^{2R_0+2R_1}(1+2t^2+2 t^4)+\ldots
\label{Ys}
\eeqa
There
are two particularly interesting classes of diagrams entering in
(\ref{Ys}). The first one will be associated to the contribution
of regular instantons and can be found by restricting to those
Young tableaux transforming in the regular representation of
$\Z_2$
\beqa Y_{\rm reg}^{\Z_2} &=&\bullet
+\left(\tinyyoung{\bullet\hfil}+\tinyyoung{\hfil,\times}\right)
+\left(\tinyyoung{\bullet\hfil\bullet\hfil}+\tinyyoung{\hfil,\bullet\bullet\hfil}
+\tinyyoung{\bullet\hfil,\hfil\times}
+\tinyyoung{\hfil,\times,\bullet\hfil}+\tinyyoung{\hfil,\times,\hfil,\times}
\right)+\ldots\nn\\
&=& 1+  (1+t^2) q^2\, r^{R_0+R_1} + q^4 r^{2R_0+2R_1}(1+2t^2+2
t^4)+\ldots
\label{Yr}
\eeqa
This is the same Poincar\'{e}
polynomial one would have obtained by studying the symmetric
product $S^k(R^4/\Z_2)$ \cite{Dijkgraaf:1996xw}.
The equivalence between the Poincar\'{e} polynomial of regular instantons
and that of symmetric products holds for all the group $\Z_p$ we have studied
\footnote{We believe that this is always the case but we don't have
a proof of this. In this paper we also report
on the $\Z_3$ case. We have also studied other cases with a Mathematica code.}.
 In mathematical terms this translates into the relation $(\C^{[k]})^\Gamma=(\C/\Gamma)^{[k]}$
 with $M^{[k]}$ the Hilbert scheme of $k$ points on $M$.
More generally
\beqa
Y_{\rm reg}^{\Z_p}
 &=&\sum_k \,q_{\rm reg}^{k}   Y( S_k(R^4/\Z_p))\nn\\
& =&\prod_{m=1}^\infty {1\over (1-q_{\rm reg}^{m}\, t^{2 m-2})(1-q_{\rm reg}^{  m}\, t^{2 m})^{p-1} }\label{Yr1}
\eeqa
with the parameter $q_{\rm reg}\equiv q^p\, r^{\rm Reg}$ tracing the number of instanton
 in the regular representation $R_{\rm reg}\equiv \sum_q R_q$.
The second class is associated to
fractional instantons and corresponds to tableaux with no
$\Gamma$-invariant boxes (i.e with no bullets or crosses)
\beqa
Y_{\rm frac}^{\Z_2} &=&\bullet
+\tinyyoung{\hfil}
 +\tinyyoung{\hfil,\hfil\hfil}+\tinyyoung{\hfil,\hfil\hfil,\hfil\hfil\hfil}
+\ldots\nn\\
&=& 1+ q \,r^{R_0}+q^3  r^{R_0+2 R_1}+ q^6 r^{4 R_0+2 R_1} +\ldots
\label{Yf} \eeqa We call this class fractional to emphasize that
this contributions carry no moduli (no bullets or crosses) i.e. they correspond to
disjoint points (zero dimensional surfaces) in the moduli space.
It is easy to find a closed form for this generating function. Let
us consider separately the two cases in which the number of boxes
in the first row are even or odd. In the former case the number of
boxes in the first row is $2k$ with $k=0, 1, \ldots$ and the
number of boxes transforming according to the $R_0$, $R_1$
representations is $k^2$ and $k^2+k$ respectively. In the latter
case, if the number of boxes is $2k-1$ with $k\in \Z_+$ the number
of boxes transforming according to the $R_0$, $R_1$
representations is $k^2$ and $k^2-k$ respectively. Then
\beqa
Y_{\rm frac}^{\Z_2} &=& 1+\sum_{k=1}^\infty \left(
q^{k(2k-1)}r^{k^2 R_0+(k^2-k) R_1} +q^{k(2k+1)}r^{k^2 R_0+(k^2+k)
R_1}\right)\nn\\
&=&\sum_{k=-\infty}^\infty
\left(qr^{R_1}\right)^k\left( q^2r^{R_0+R_1}\right)^{k^2}=\theta_3 (z_1|q^2_{\rm reg})
\label{Yf1}
\eeqa
with $e^{2\pi i z_1}\equiv q r^{R_1}$ and $q_{\rm reg}\equiv q^p r^{R_{\rm reg}}$ with $R_{\rm reg}={\sum_q R_q}$.
Conventions for theta functions are explained in Appendix \ref{theta}.
Each power of $e^{2\pi i z_1}$ correspond to an unpaired fractional instanton in the
$R_1$-representation of $\Z_2$.
Remarkably the full instanton  character $Y$ can be
written as the product of the regular $Y_{\rm reg}$ and fractional
$Y_{\rm frac}$ contributions
\beq Y=Y^{\Z_2}_{\rm reg}Y^{\Z_2}_{\rm frac}
\label{magic}
\eeq
as can be explicitly checked by taking the product
of (\ref{Yr1}) and (\ref{Yf1}) and comparing against (\ref{Ys}).
 The Poincar\'{e} polynomial can then be written as
 \beq
P_t=\sum_{m,k,{\cal R}} b_m({\cal M}_{k,{\cal R}}) \, q^k\, t^m\, r^{\cal R}=
   \prod_{n=1}^\infty{  (1-q_{\rm reg}^{2 n}) (1+q\,r^{R_1} q_{\rm reg}^{2 n-1})(1+ (qr^{R_1})^{-1} q_{\rm reg}^{2 n-1})
  \over (1-q_{\rm reg}^n t^{2n-2})(1-q_{\rm reg}^n t^{2n})}
\label{magic1}
\eeq

Similar relations hold for $\Z_3$. Now
\beqa Y &=&\bullet
+\tinyyoung{\hfil}+\left(\tinyyoung{\hfil\hfil}+\tinyyoung{\hfil,\hfil}\right)
+\left(\tinyyoung{\bullet\hfil\hfil}+\tinyyoung{\hfil,\bullet\hfil}+\tinyyoung{\hfil,\hfil,\times}\right)
+\left(\tinyyoung{\hfil\bullet\hfil\hfil}+\tinyyoung{\hfil,\hfil\hfil\hfil}
+\tinyyoung{\hfil\hfil,\bullet\hfil}
+\tinyyoung{\hfil,\hfil,\hfil\hfil}+\tinyyoung{\hfil,\hfil,\times,\hfil}
\right)+\ldots
\nn\\
&=& 1+ q \,r^{R_0}+ q^2\, (r^{R_0+R_1}+r^{R_0+R_2})+q^3
r^{R_0+R_1+R_2}(1+2 t^2)\nn\\
&&\quad\quad ~~~~~~~~~+ q^4 (r^{R_0+2R_1+R_2}
+r^{R_0+R_1+2R_2}+(1+2t^2)r^{2R_0+R_1+R_2})\ldots
\label{Ys3}
\eeqa
and
\beqa Y_{\rm reg}^{\Z_3} &=&\bullet
+\left(\tinyyoung{\bullet\hfil\hfil}+\tinyyoung{\hfil,\bullet\hfil}+\tinyyoung{\hfil,\hfil,\times}\right)
+\ldots\nn\\
&=& 1+q^3 \,r^{R_0+R_1+R_2} (1+2 t^2)+\ldots \nn\\
Y_{\rm frac}^{\Z_3}
&=&\bullet+\tinyyoung{\hfil}+\left(\tinyyoung{\hfil\hfil}+\tinyyoung{\hfil,\hfil}\right)+
+\left(\tinyyoung{\hfil,\hfil\hfil\hfil}+\tinyyoung{\hfil,\hfil,\hfil\hfil}\right)
+\tinyyoung{\hfil,\hfil,\hfil\hfil\hfil}+\ldots\label{Yf3}\\
&=& 1+q \,r^{R_0}+q^2\, (r^{R_0+R_1}+r^{R_0+R_2})+ q^4
\,(r^{R_0+R_1+2R_2}+r^{R_0+2R_1+R_2})
+\ldots\nn
\eeqa
Again the relation (\ref{magic}) is verified. The extension of this algorithm to the case of $U(n)$
is a straightforward matter.

\subsection{The ${\cal N}=4$ partition function on the Eguchi-Hanson manifold for gauge group $SU(2)$}

In this last section of this chapter, we compute the Euler characteristic for the moduli space
of gauge connections with half integer winding number on the Eguchi-Hanson manifold which is
the simplest ALE variety. The Euler characteristic of the multi-instanton moduli space
is the ${\cal N}=4$ partition function of the
theory. Taking the limit of zero mass in the partition function (\ref{generalsdetn=2*}) we in fact
find that the  ${\cal N}=4$ partition function is given by the sum over all critical points i.e.
the Euler characteristic. With respect to the $\R^4$ case \cite{Bruzzo:2002xf} the novelty is that
for the $SU(2)$ Eguchi-Hanson case
we have to consider only those critical points obeying to the conditions
\beqa
 q_1=q_2=1:&& \qquad k=(k_0,k_0+1)\qquad   n=(0,2)\nn\\
q_1=q_2=0:&& \qquad k=(k_0,k_0)\qquad      n=(2,0)~~\nn
\eeqa
These conditions force us to count all the pairs of
Young tableaux with an odd(even) number of boxes belonging to the classes $k_0 R_{\rm reg}+R_1$($k_0 R_{\rm reg}$)
with the first box transforming according to the $R_1$($R_0$) representation
of $\Z_p$. The starting point is the $U(2)$ Euler polynomial $Y_{U(2)}=(Y^{\Z_2}_{\rm reg}Y^{\Z_2}_{\rm frac})^2|_{t=1}$
given by  (\ref{Yr1},\ref{Yf1}) i.e. before
imposing the $c_1=0$ condition.
  Singling out the classes $k_0 R_{\rm reg}$
and $k_0 R_{\rm reg}+R_1$ in this formula it is now easy.
Unpaired instantons come only from
\beq
Y_{\rm frac}^{\Z_2}=\sum_{k=-\infty}^\infty
\left(qr^{R_{1+q_1}}\right)^k\, q_{\rm reg}^{k^2}
\label{tetainter}
\eeq
where we have modified the subscript of the term $R_1$ to generalize (\ref{Yf1})
which was found in the case $q_1=0$\footnote{We remind the reader that the subscripts of the representations
are always understood modulo $p$.}.
Then
\beq
Y_{U(2)}=(Y^{\Z_2}_{\rm reg})^2\sum_{k_1,k_2=-\infty}^\infty
\left(qr^{R_{1+q_1}}\right)^{k_1}\left(qr^{R_{1+q_2}}\right)^{k_2}\left( q_{\rm reg}\right)^{k_1^2+k_2^2}.
\label{tetainter1}
\eeq
The factors $(qr^{R_{1+q_1}})^{k_1}$ and $(qr^{R_{1+q_2}})^{k_2}$ in (\ref{tetainter}) are the only
potential sources of asymmetry between the $R_0$ and $R_1$ representations.
The classes $k_0 R_{\rm reg}$ (with $q_1=q_2=0$) and $k_0 R_{\rm reg}+R_1$ (with $q_1=q_2=1$)
are extracted from (\ref{tetainter1}) by picking
up the terms with $k_1=-k_2$ and $k_1=-k_2-1$ and specializing to $r=1$
\beqa
Y^{SU(2)}_{\rm even}&=&q_{\rm reg}^{-{1\over 6}}\, (Y^{\Z_2}_{\rm reg})^2\sum_{k=-\infty}^\infty\, q_{\rm reg}^{2k^2}=
{ \theta_3(0|q_{\rm reg}^4) \over \eta(q_{\rm reg})^4} \nn\\
Y^{SU(2)}_{\rm odd}&=&q_{\rm reg}^{-{1\over 6}}\,(Y^{\Z_2}_{\rm reg})^2\,\sum_{k=-\infty}^\infty  q_{\rm reg}^{{4\over 2}(k-{1\over 2})^2}
=   {\theta_2(0|q_{\rm reg}^4)\over \eta(q_{\rm reg})^4}
\label{tetainter2}
\eeqa
The factor $q_{\rm reg}^{-{1\over 6}}$ (Casimir energy) has been included as in \cite{Vafa:1994tf}
in order to reconstruct the modular functions.
See the Appendix and (\ref{teta3}) for our notation.
The partition function of ${\cal N}=4$ is then
\beq
{\cal Z}_{{\cal N}=4}=q^{1\over 6} Y^{SU(2)}=
{1\over\eta^4(q_{\rm reg})}\left[\theta_3(0|q_{\rm reg}^4)+\theta_2(0|q_{\rm reg}^4)\right]=
{\theta_3(0|\tau)\over\eta^4(\tau)}
\label{partfuncn=4}
\eeq
Notice that the number of boxes in the diagrams is one half the Chern class ({\ref{c2}).
Therefore the correct expansion parameter for the  $\R^4/\Z_2$ case is $q_{\rm reg}=q^2$.
(\ref{partfuncn=4}) is a modular form as expected from the $SL(2,\Z)$ duality
of ${\cal N}=4$  \cite{Vafa:1994tf}.

\section{The prepotential}
\setcounter{equation}{0}

Finally we derive the multi instanton partition functions and
prepotentials describing the ${\cal N}=2$ low energy physics on the ALE
manifold \footnote{In this chapter we will give the lowest terms
in the expansion of the prepotential using an analytical method
for pedagogical purposes. To check highest order terms, this way
is not practical and we have written a Mathematica code.}.
For simplicity we take
$\epsilon_2=-\epsilon_1=\hbar$.

The ALE projection is defined by omitting the
eigenvalues in (\ref{generalsdetn=2*}) that
are not invariant under the $\Z_p$ projections
\beq
 \Z_p: \hbar\to \hbar+{2\pi\over p}\qquad a_\alpha\to a_\alpha+
{2\pi q_\alpha\over p}
\label{zps1}
\eeq

 \subsection*{ $SU(2)$ gauge theory on $R^4/\Z_2$ }

For concreteness we consider $SU(2)$ gauge theory on $\R^4/\Z_2$.
The
condition of vanishing of the first Chern class  (\ref{c10}), $c_1=0$, is only satisfied by the
instanton configurations \cite{Bianchi:1996zj}
$k=(k_0,k_0), n=(2,0) (q_1=q_2=0)$ and $k=(k_0,k_0+1),  n=(0,2) (q_1=q_2=1)$.
According to (\ref{c2}) they correspond to
bundles with second Chern classes  $c_2=k_0$ and $c_2^\prime=\frac{1}{2}(2 k_0+1)$ respectively.

The instanton partition function on the orbifold, follows from that in flat space
after imposing invariance under (\ref{zps1}).
Thus only the eigenvalues (appearing as factors in (\ref{fst}))
invariant under (\ref{zps1}) must be taken into account.
In addition the ALE partition function is defined by
$$
{\cal Z}(q)=\sum_k {\cal Z}_k q^{{k\over 2}}
$$
since $c_2=\ft12 k$.

Furthermore let us remark that (\ref{zr41}) only depend on $a_{\alpha\beta}$.
Since for instanton configurations with $c_1=0$ one has $q_\alpha=q_\beta$,
under (\ref{zps1}) we get $\delta a_{\alpha\beta}=\pi(q_\alpha-q_\beta)=0$.
We now have to examine the behavior under $\hbar\to \hbar+\pi$. Only functions
of the type $f(n\hbar), T_\alpha(n\hbar)$ with $n$ even survive this projection.
Collecting these type of terms in (\ref{zr41}) and setting $a_1=-a_2=a$ one finds
\beqa
Z_{\tinyyoung{\hfil}}&=& 2\left(1-{m^2\over 4a^2}\right)\label{zale1}\\
Z_{\tinyyoung{\hfil\hfil}}&=&Z_{\tinyyoung{\hfil,\hfil}}=2
\left(1-{m^2\over (2\hbar)^2}\right)\left(1-{m^2\over 4 a^2}\right)\label{zale2}
\eeqa
The configuration $Z_{\tinyyoung{\hfil},\tinyyoung{\hfil}}$ does not contribute, since it has $c_1\neq 0$.
Computing also terms which come from Young tableaux with three and four boxes we finally get
\beqa
&&Z(q)=1+{(4a^2-m^2)\over 2a^2}q^{1\over 2}+{(4a^2-m^2)(4\hbar^2-m^2)\over 4a^2\hbar^2}q\nn\\
&&+{m^2 a^2(4a^2-m^2)+\hbar^2(128 a^6-64m^2a^4+8m^4a^2-3m^6)\over 16a^6\hbar^2}q^{3\over 2}\nn\\
&&+\biggl[{4m^4 a^4(-4a^2+m^2)
+ \hbar^2a^2(-704a^6m^2+352a^4m^4-120a^2m^6+5m^8)\over 128a^8\hbar^4} \nn\\+
&&{\hbar^4(2048a^8-1152m^2a^6+640m^4a^4-152a^2m^6+9m^8)\over 128a^8\hbar^4}\biggl]q^2+\ldots
\label{zn=2*}
\eeqa
>From (\ref{zn=2*}) we can extract the partition function for the pure ${\cal N}=2$ case
by taking the limit $m\to\infty, m^4q=\Lambda$
\beq
Z(\Lambda)=1-\frac{\Lambda^{1\over 2}}{2 \, {a}^{2}}+\frac{{\Lambda}}{4 \,
{a}^{2} \, {\hbar}^{2}}- {\Lambda}^{3\over 2}
\frac{(a^2+3\hbar^2)}{8 \hbar^2 a^6}
 + {\Lambda}^{2}\frac{\left ( 4 \, {a}^{4}+5 \, {a}^{2} \, {\hbar}^{2}+9 \,
{\hbar}^{4}\right ) }{128\, {a}^{8} \, {\hbar}^{4}}+\ldots
\label{zn=2} \eeq

The presence of half-integer powers of $q$ makes $Z(q)$ double-valued in the
complex plane, i.e. $Z(q)$ is not invariant under
$q^{1\over 2}\to -q^{1\over 2}$ or $\tau\to \tau+1$.
 Out of $Z(q)$ we can build two objects with definite transformation
 properties under $\tau\to \tau+1$:
 \beqa
{\cal F}_{\rm even}(q,m,\hbar)&=& \ft12\hbar^2\, \left[\ln\,Z(q^{1\over 2}) +\ln Z(-q^{1\over 2})\right]\nn\\
 {\cal F}_{\rm odd}(q,m,\hbar)&=& -\ft12\, \left[\ln\,Z(q^{1\over 2}) -\ln Z(-q^{1\over 2})\right]
\label{prep2}
\eeqa
The two terms correspond to contributions with integer and half-integer Chern class respectively.
Notice the different normalization of the two pieces.
The ${1\over \hbar^2}$ volume factor in front of the odd term has been projected out by the
$\Z_2$ orbifold group action.
Notwithstanding the odd term is now regular in the limit $\hbar\to 0$.
The quadratic divergence is in fact connected to the breaking of
translational invariance\footnote{This translational symmetry of the lagrangian in the moduli space
follows from the translational invariance of the instanton solution in Euclidean space time.
In \cite{Nekrasov:2002qd} it was suggested to deform the original
lagrangian under $T_{\epsilon_{1,2}}$ to allow for a localization with isolated critical points.}
in the lagrangian of the theory due to the rotations $T_{\epsilon_{1,2}}$. The lagrangian
invariant under  the infinitesimal action of $T_{\epsilon_{1,2}}$
explicitly contains the modulus which gives the position of the center of the instanton.
The integration over this modulus is responsible for the above mentioned quadratic divergence.
The $\Z_2$ invariant instanton is clearly centered in zero and there is no modulus for its center.
For ${\cal N}=2^*$ one finds
\beqa
{\cal F}_{\rm even} (q) &=& \left({m^4-4m^2a^2\over  4a^2}\right)q+
\frac{m^2}{128a^6}(5m^6-48m^4a^2+96m^2a^4-192a^6)q^2\nn\\
&&+\hbar^2\left[\left({16a^4-m^4\over  8a^4}\right)q
+\left({-17m^8+72m^6a^2-128m^2a^6+512a^8\over 128 a^8}\right)q^2\right]+\ldots \nn\\
{\cal F}_{\rm odd} (q) &=&\left({m^2-4a^2\over 2a^2}\right)q^{1\over 2}
-\left({32a^6-24m^2a^4+24m^4a^2-5m^6\over 12 a^6}\right)q^{3\over 2}\nn\\
&&+\left({48m^2a^4-32m^4a^2+5m^6\over 8 a^8}\right)q^{3\over 2}
+\ldots
\eeqa
Once again  the pure ${\cal N}=2$ limit is recovered in the limit $m\to\infty, m^4q=\Lambda$
\beqa
{\cal F}_{\rm even} (\Lambda) &=& \left(\frac{1}{4}{\Lambda\over  a^2}+
\frac{5}{128}  {\Lambda^2\over a^6}+\frac{3}{128}  {\Lambda^3\over a^{10}}\right)
-\hbar^2\left(\frac{1}{8}  {\Lambda\over a^4}+\frac{17}{128}  {\Lambda^2\over a^8}
+\frac{13}{64}  {\Lambda^3\over a^{12}}\right)+\ldots \nn\\
{\cal F}_{\rm odd} (\Lambda) &=&\left(\frac{1}{2}{\Lambda^{1\over 2}\over a^2}
+\frac{5}{12}{\Lambda^{3\over 2}\over a^6}+\frac{207}{320}{\Lambda^{5\over 2}\over a^{10}}\right)
+\hbar^2\left(\frac{5}{8}{\Lambda^{3\over 2}\over a^8}+
\frac{269}{64}{\Lambda^{5\over 2}\over a^{12}}\right)
+\ldots
\eeqa
The $k=1/2,\hbar=0$ term matches (4.14) in \cite{Bellisai:1997ap}\footnote{In that reference
the correlator $Tr\phi^2$ is computed. For the lowest winding number this is the same as the prepotential.}.
For completeness we have also given the first gravitational correction to the prepotential.

\section*{Acknowledgements}
The authors want to thank A. Maffei for very enlightening discussions on the properties of
moduli spaces of gauge connections on ALE manifolds and  R.Flume for collaboration in
the early stage of this work.
R.P. have been partially supported by the Volkswagen foundation
of Germany and he also would like to thank I.N.F.N. for supporting
a visit to the University of Tor Vergata.
This work was supported in part by the EC contract
HPRN-CT-2000-00122, the EC contract HPRN-CT-2000-00148,
the EC contract HPRN-CT-2000-00131,
the MIUR-COFIN contract 2003-023852, the NATO
contract PST.CLG.978785 and the INTAS contracts 03-51-6346 and 00-561.

\begin{appendix}

\section{Theta functions}
\label{theta}

The conventions for theta functions in the text are:
\beqa
\theta[^a_b](z|q)&=&\sum_{n\in Z}q^{{1\over 2}\left(n-a\right)^2}
e^{2\pi i\left(z-b\right)\left(n-a\right)}\nn\\
\eta(q)&=&q^{\frac{1}{24}}\prod_{n=1}^\infty(1-q^n)
\label{teta3}
\eeqa
where  $\theta_1=\theta [^{1\over 2}_{1\over 2}]$, $\theta_2=\theta[^{1\over 2}_0]$,
$\theta_3=\theta[^0_0]$, $\theta_4=\theta[^0_{1\over 2}]$.

\section{Cohomology of the ALE space}

As an illustration here we compute the homologies of the ALE space itself.
This is a well known result (see \cite{Eguchi:jx} for a review). Here we will derive it  using the methods
of subsection \ref{gaugesec}: the ALE space can be described in terms of the
non commutative ADHM formalism
for the $U(1)$ case.
In fact we have seen in subsection \ref{kronnakajima}
that the matrices which describe the ALE space are subject to the constraint
(\ref{duetredicibb}) and to the $\Gamma$ invariant condition (\ref{gamminva}).
If in the ADHM construction we described in subsection \ref{gaugesec} we take
the $U(1)$ case (i.e. the space $W$ becomes a $k$ dimensional vector), we can
set $J=0$ and the constraints $f_\C=0, f_\R=\zeta$ coincide with (\ref{duetredicibb})\footnote{The deformation
appearing in (\ref{duetredicibb}) is subject to the condition $\sum_{i=1}^{p-1} \zeta_i=0$
while the $\zeta$ in $f_\R=\zeta$ is unconstrained.
The two conditions agree if we take into
account the term $II^\dagger$ appearing in $f_\R$.}
 Moreover choosing $k=p$ in the regular representation the projection on ADHM moduli
 space matches that in (\ref{gamminva}).
 We can then compute the homologies of an ALE space using the character $\chi$ in
(\ref{tracef}).
  The $\Gamma$ invariance requires in this case $h(s)+v(s)+1=p$.
How many boxes in the set of diagrams with $\vert Y\vert=\vert\Gamma\vert$ obey this condition?
The answer is given in Fig.\ref{figura3}.
\begin{figure}[ht]
\label{figura3}
\setlength{\unitlength}{2mm}
\begin{center}
\begin{picture}(70,25)(0,-15)

\put(0,0){\framebox(3,3){$\bullet$}}\put(3,0){\framebox(3,3)}
\put(6.4,1){$\cdots$}\put(9.6,0){\framebox(3,3)}
\put(18,0){\framebox(3,3){$\bullet$}}\put(18,-3){\framebox(3,3)}\put(21,0){\framebox(3,3)}
\put(24.4,1){$\cdots$}\put(27.6,0){\framebox(3,3)}
\put(36,0){\framebox(3,3){$\bullet$}}\put(36,-3){\framebox(3,3)}
\put(36,-6){\framebox(3,3)}\put(39,0){\framebox(3,3)}
\put(42.4,1){$\cdots$}\put(45.6,0){\framebox(3,3)}
\put(54,1){$\cdots$}
\put(63,0){\framebox(3,3){$\times$}}\put(63,-3){\framebox(3,3)}\put(63,-6){\framebox(3,3)}
\put(64,-9){$\vdots$}\put(63,-12.6){\framebox(3,3)}

\end{picture}
\caption{Subset of diagrams with $p$ boxes satisfying $h(s)+v(s)+1=p$.}
\end{center}
\end{figure}
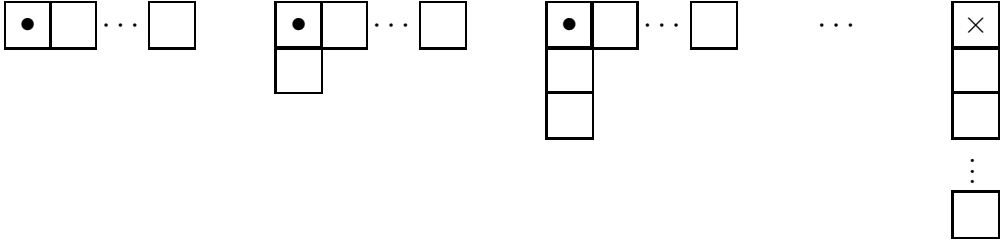

There we draw the only diagrams in which at least
one box satisfies the $\Gamma$ invariance condition. As before "$\bullet$"
refers to $\Gamma$-invariant boxes with $h(s)>0$ while "$\times$" stands for $h(s)=0$ .
Counting the number of bullets one finds the homology
\begin{itemize}
\item $b_0=1$ since the box denoted by a $\times$ in the last diagram
from the left in  Fig.\ref{figura3} satisfies the $\Gamma$ invariance condition but has $h(s)=0$
\item $b_2=p-1$ since there are $p-1$ diagrams with one negative eigenvalue, given by the boxes
with a $\bullet$.
\end{itemize}

\end{appendix}

\end{document}